\documentclass[a4paper,11pt]{article}
\usepackage{pos}
\usepackage{comment}
\usepackage{tikz}
\usepackage{tikzscale}
\usepackage{neuralnetwork}
\usepackage{
url,
graphicx,
epsfig,
graphics,
subfigure,
psfrag,
amsmath,
epstopdf,
enumerate,
lineno,
hyperref}
\usepackage{xcolor}
\usepackage{marginnote}

\usepackage[normalem]{ulem}

\title{Exploring Generative Networks for Manifolds with Non-Trivial Topology}

\author*[a]{Shi-Yang Chen}
\author[a]{Gert Aarts}
\author[b]{Biagio Lucini}
\affiliation[a]{ Department of Physics, Swansea University, SA2 8PP, Swansea, United Kingdom}
\affiliation[b]{Department of Mathematics, Swansea University (Bay Campus),
  Swansea, SA1 8EN, United Kingdom}

\emailAdd{2222895@swansea.ac.uk}
\emailAdd{g.aarts@swansea.ac.uk}
\emailAdd{b.lucini@swansea.ac.uk}

\abstract{The expressive power of neural networks in modelling non-trivial distributions can in principle be exploited to bypass topological freezing and critical slowing down in simulations of lattice field theories. Some popular approaches are unable to sample correctly non-trivial topology, which may lead to some classes of configurations not being generated. In this contribution, we present a novel generative method inspired by a model previously introduced in the ML community (GFlowNets). We demonstrate its efficiency at exploring ergodically configuration manifolds with non-trivial topology through applications such as triple ring models and two-dimensional lattice scalar field theory.}

\FullConference{%
 The 41th International Symposium on Lattice Field Theory, LATTICE2024
  28th July -3rd August, 2024
  the University of Liverpool, United Kingdom
}

\begin{document}
\maketitle

\section{Introduction}

Neural networks can be employed to represent numerically systems with many strongly-interacting degrees of freedom, such as quarks and gluons in Quantum Chromodynamics (QCD).
Currently, the main tool for understanding QCD is the Monte Carlo Markov Chain (MCMC) method. However, non-trivial challenges can affect MCMC simulations, including critical slowing down \cite{Wolff:1989wq} and topological freezing \cite{DelDebbio:2004xh,Schaefer:2010hu}. 
The very large number of neurons in modern neural network architectures could improve the understanding of statistical field theories and at the same time help to model complex distributions; see, e.g., the reviews~\cite{Cranmer:2023xbe,Kanwar:2024ujc,Aarts:2025gyp}.


The most widely used neural network architectures can be classified into two types based on how they learn. One type of architecture is trained on a pre-existing data set. This is commonly referred to as supervised learning. Examples with applications in lattice field theory include Generative Adversarial Networks \cite{Pawlowski:2018qxs,Wang:2020hji} and Diffusion Models (DMs) \cite{Wang:2023exq,Wang:2023sry,Zhu:2024kiu,Aarts:2024rsl}. 
In contrast, the second type learns directly, without any need for training data, on a reward function or a probability density, as in Reinforcement Learning, Normalising Flow (NFs)
\cite{rezende2015variational,Noe:2019,Albergo:2019eim,Nicoli:2019gun,Kanwar:2020xzo,Nicoli:2020njz,Nicoli:2023qsl} and variations thereof, such as Continuous Normalizing Flow~\cite{Chen:2018,deHaan:2021erb, Gerdes:2022eve,Caselle:2023mvh}. This is referred to as unsupervised learning. 

Ref.~\cite{wu2020stochastic} pointed out that NFs may suffer from non-ergodicity in the presence of topological constraints: since the network learns an invertible map between the prior and the target distribution, issues appear when the prior distribution is unimodal, e.g.\ a Gaussian, but the target distribution is bi- or multimodal. Ref.~\cite{wu2020stochastic} contains some illustrative examples for two-dimensional images and proposes a modified version called Stochastic Normalizing Flow, which was subsequently used in lattice field theory \cite{Caselle:2022acb,Caselle:2024ent}. Ref.~\cite{Nicoli:2023qsl} contains a detailed study of mode collapse in lattice field theory when using NFs, demonstrating issues arising with modelling multimodal distributions. 

In this contribution we focus on approaches where the model is trained directly on the target distribution, i.e., unsupervised learning models. Besides NFs, we introduce unsupervised DMs, which should not be confused with the DMs studied in Refs.~\cite{Wang:2023exq,Wang:2023sry,Zhu:2024kiu,Aarts:2024rsl}. 
We start by investigating the ability of NFs and unsupervised DMs to learn global features for given probability distributions. A triple-ring model is used as a target manifold to study their performance. We demonstrate that topology plays an essential role in the generation process and determine the global features of the output manifold. To circumvent the problem with topology, we propose a new structure based on DMs and verify its effectiveness for the triple-ring model. Finally, the magnetisation in a two-dimensional $\phi^4$ lattice theory is studied via the above neural network architectures.
Our method is inspired by an approach previously introduced in the ML community, called GFlowNets \cite{DBLP:journals/corr/abs-2106-04399, DBLP:journals/corr/abs-2111-09266}.


\section{Problem description}

Normalizing flows and diffusion models are two of the most widely used machine learning generative methods in lattice field theory. NF shapes a distribution $P_{\rm target}(z_t)$ from a prior $P(z_0)$ by changing the volume element \cite{Albergo:2019eim},
\begin{equation}
     P_{\rm target}(z_t) = P_0(f(z_t)) \left|\frac{\partial f(z_t)}{\partial z_t}\right|,
    \label{variable_charge}
\end{equation}
where $f^{-1}$ is a function with learnable parameters $\Theta$, mapping an element $z_0$ from the prior distribution to the target distribution, $z_t = f^{-1}(z_0)$. 
Differentiability and reversibility from Eq.~\eqref{variable_charge} demands that NFs are a bijection, which implies that the modelled manifold from NFs is homeomorphic to the prior manifold.

Diffusion models evolve the prior distribution towards a  target distribution at some time $T$. The pathway of evolution is driven by the Langevin equation,
\begin{equation}
	z_{t+1} = z_{t} - K(z_t)D^2(z_t)dt + \sqrt{2dt}D(z_t)\eta_t, \quad\quad \langle\eta_t\rangle_\eta=0, \quad \langle\eta_s\eta_t\rangle_\eta=\delta(s-t),
 \label{langevin}
\end{equation}
where $D(z_t)$ and $K(z_t)$ the diffusion and drift estimated by neural networks with $t$ the fictitious time, and $\eta_t$ is white or Gaussian noise. The stochasticity enables DMs to generate multiple $z_t$'s, resulting in different possible paths, as in the path integral approach, similar to stochastic quantisation \cite{Wang:2023exq}.
The connectivity of the output manifold $z_{t+1}$ is verified, if the manifolds $K(z_t)$ and $D(z_t)$ are connected, since a vector space formed by multiplication or addition of two connected vector spaces is also connected. Therefore, the connectivity of the manifolds $K(z_t)$ and $D(z_t)$ will determinate the topology of the output manifold $z_{t+1}$. Due to the linear multiplication and continuous activation functions, the manifolds $K(z_t)$ and $D(z_t)$ from neural networks are connected if the input manifold $z_{t}$ is connected. 

\begin{figure}[thbp]
\centering
\includegraphics[width=0.80\textwidth]{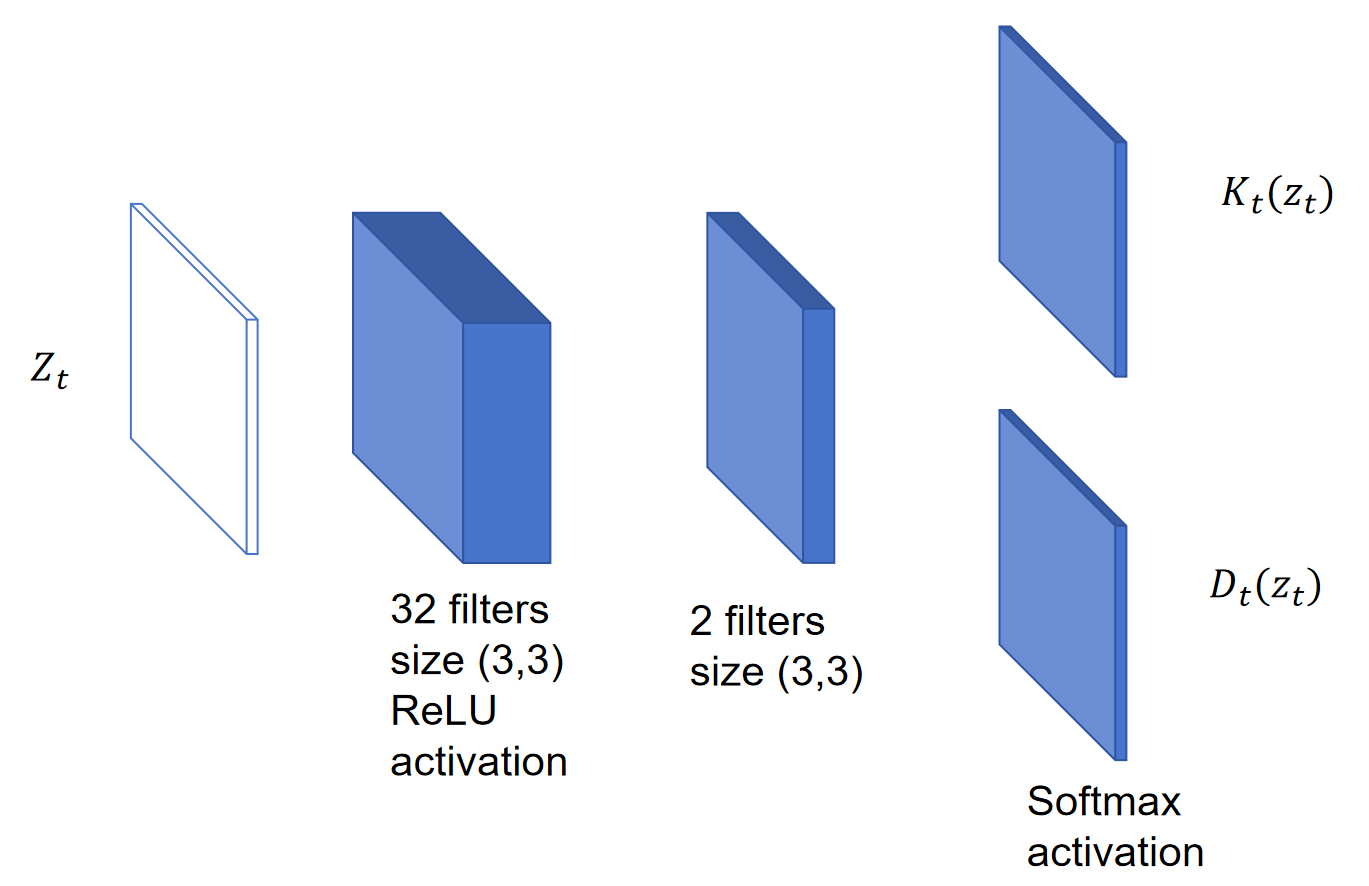}
\caption{The architecture for forward and backward processes at a fictitious time $t$.}
\label{diffusionnetwork0}
\end{figure}

To verify the above conclusion, NFs and unsupervised DMs are used to generate distributions with three rings in the two-dimensional plane (triple-ring model from now on), 
\begin{equation}
    P_{\rm rings}(r) = \exp\left(-\sum_{i=1}^3(r-\mu_i)^2/0.01 \right), \qquad r=\sqrt{x^2+y^2},
\end{equation}
with $\mu_1=0.5, \mu_2=1.5$ and $\mu_3=2.5$. The prior distribution is a Gaussian one. The NF model we use here is the one proposed in Ref.~\cite{Albergo:2021vyo}. As stated, our unsupervised DMs learn from the logarithm of the target distribution, rather than from a training dataset. 
The corresponding cost function (KL divergence) reads
\begin{equation}
	D_{\rm KL}(P(z_t||Q(z_t)) = \int P(z_t) \log \frac{P(z_t)}{Q(z_t)} dz_t,
	\label{kl_divergence}
\end{equation}
where $Q(z_t)$ the target distribution, and the learnt distribution is written as
	\begin{align}
	P(z_t) &= \frac{P(z_t, z_{t-1},\cdots, z_{1}|z_0)}{P( z_{t-1}, z_{t-2}\cdots, z_{0}|z_t)}P(z_0) \nonumber \\
			&=\frac{P_{f,t-1}(z_t|z_{t-1})P_{f,t-2}(z_{t-1}|z_{t-2})\cdots P_{f,0}(z_1|z_{0})}{P_{b,t}(z_{t-1}|z_{t})P_{b,t-1}(z_{t-2}|z_{t-1})\cdots P_{b,1}(z_0|z_{1})}P(z_0),
	\label{Bayes}
	\end{align}
with the assumption that $P_{f,t - 1}(z_t|z_{t-1}, z_{t-2})=P_{f,t-1}(z_t|z_{t-1})$, and analogously for $P_{b,t}$. The forward process and corresponding distribution $P_{f,t}(z_{t+1}|z_{t})$ from $t$ to $t+1$ is estimated via Eq.~\eqref{langevin}. Similarly, $P_{b,t}(z_{t-1}|z_{t})$ is the backward process from $t$ to $t-1$. The forward and backward processes are estimated by a series of neural networks -- see Fig.~\ref{diffusionnetwork0} -- with an argument $z_{t}$ and two output drift $K(z_t)$ and diffusion $D(z_t)$. The networks for the forward and backward processes share the same structure, but differ in terms of weights and biases. For a certain $t$, the argument $z_{t}$ is processed by a 2D-convolution layer with 32 filters. After a non-linear mapping by $\mathbf{ReLU}$, it flows into another 2D-convolution layer with 2 filters. The processed data flow is split into two parts. One is the drift and the other one, after applying the $\mathbf{Softplus}$ activation function mapping, is the diffusion.

\begin{figure}[thbp]
\centering
\subfigure[Target distribution.]{\label{Target}
\includegraphics[width=0.4\textwidth]{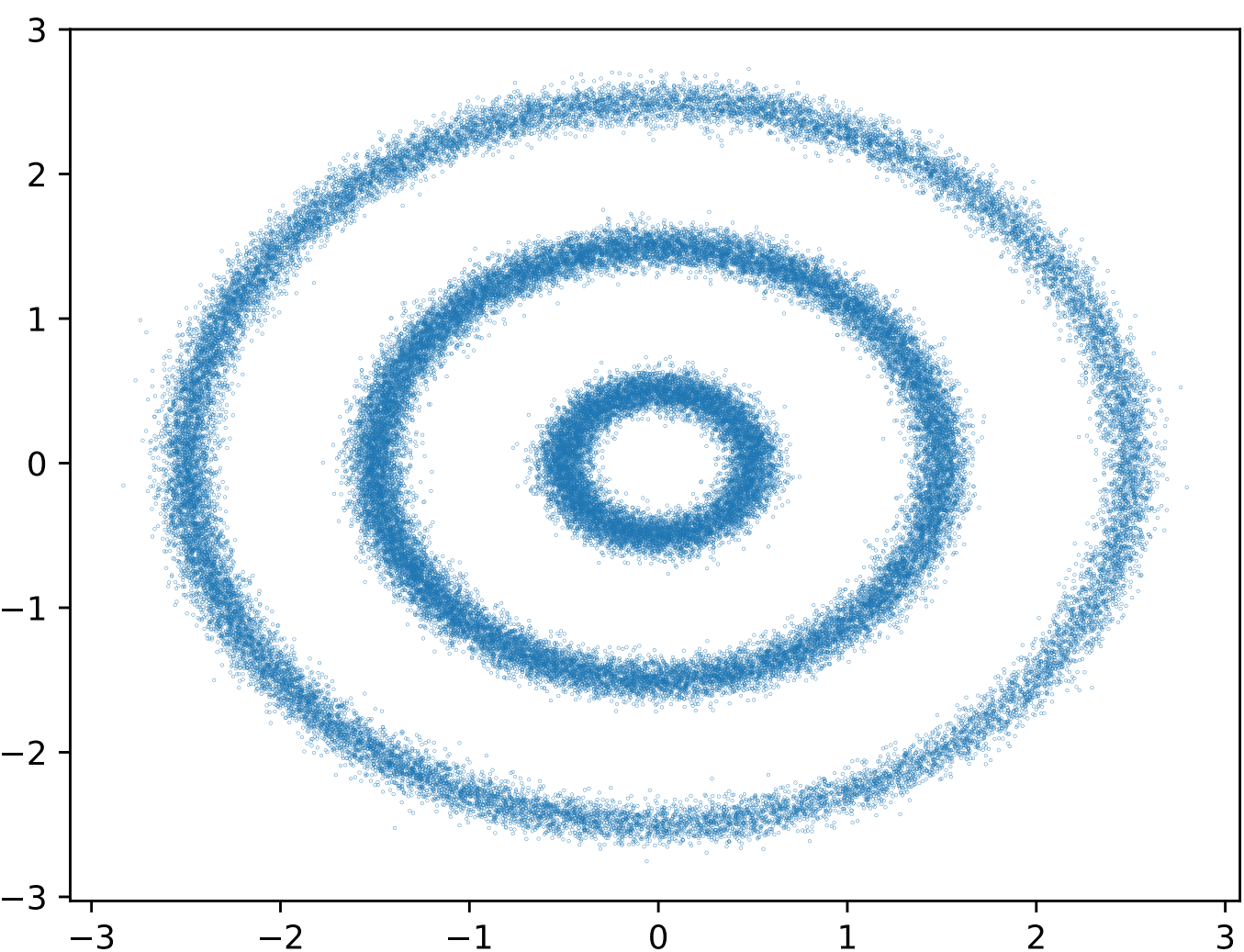}}
\;\;\;\;\;\;\;\;
\subfigure[Prior distribution.]{\label{Prior}
\includegraphics[width=0.4\textwidth]{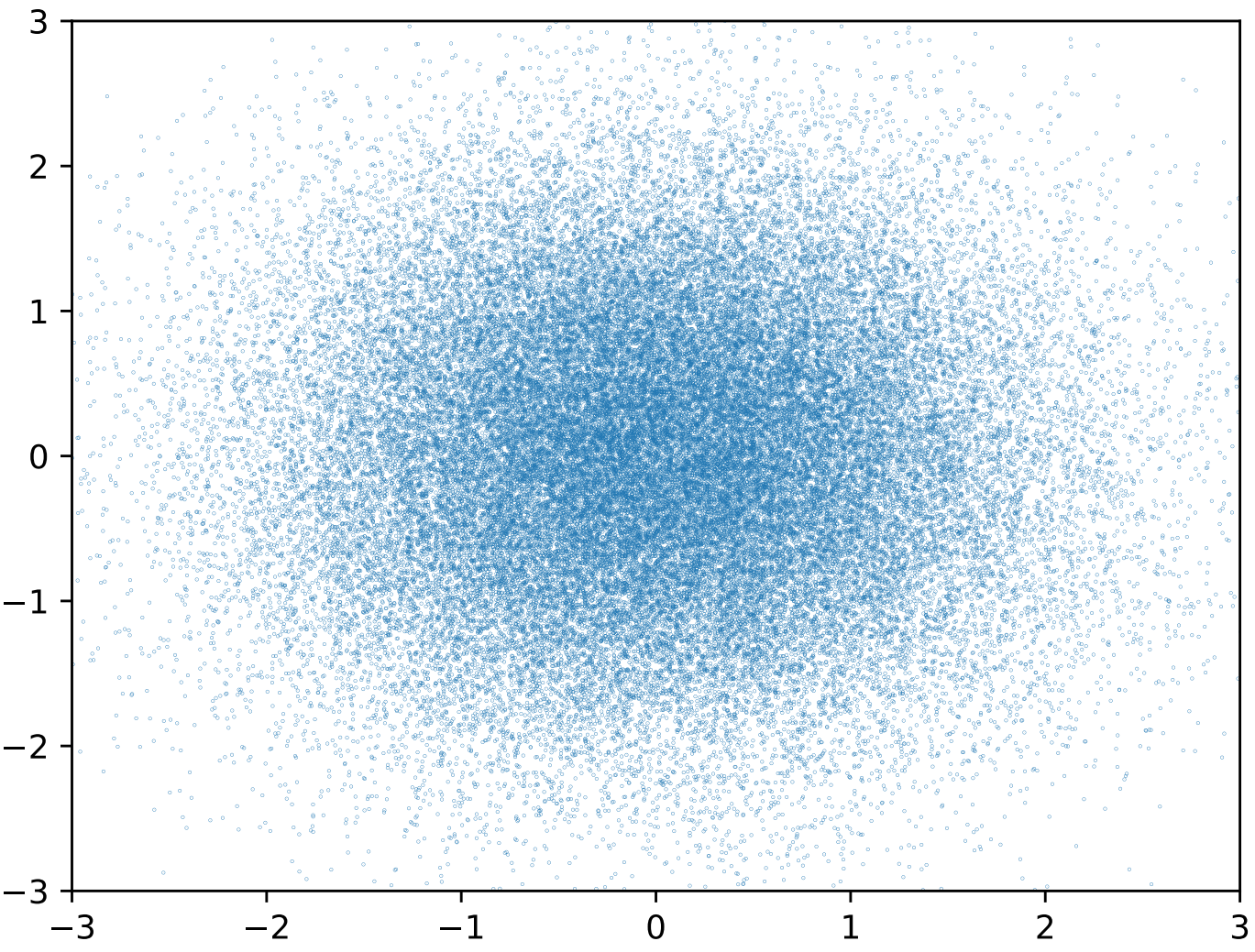}}\\
\subfigure[Distribution generated using normalizing flow.]{\label{NFring}
\includegraphics[width=0.4\textwidth]{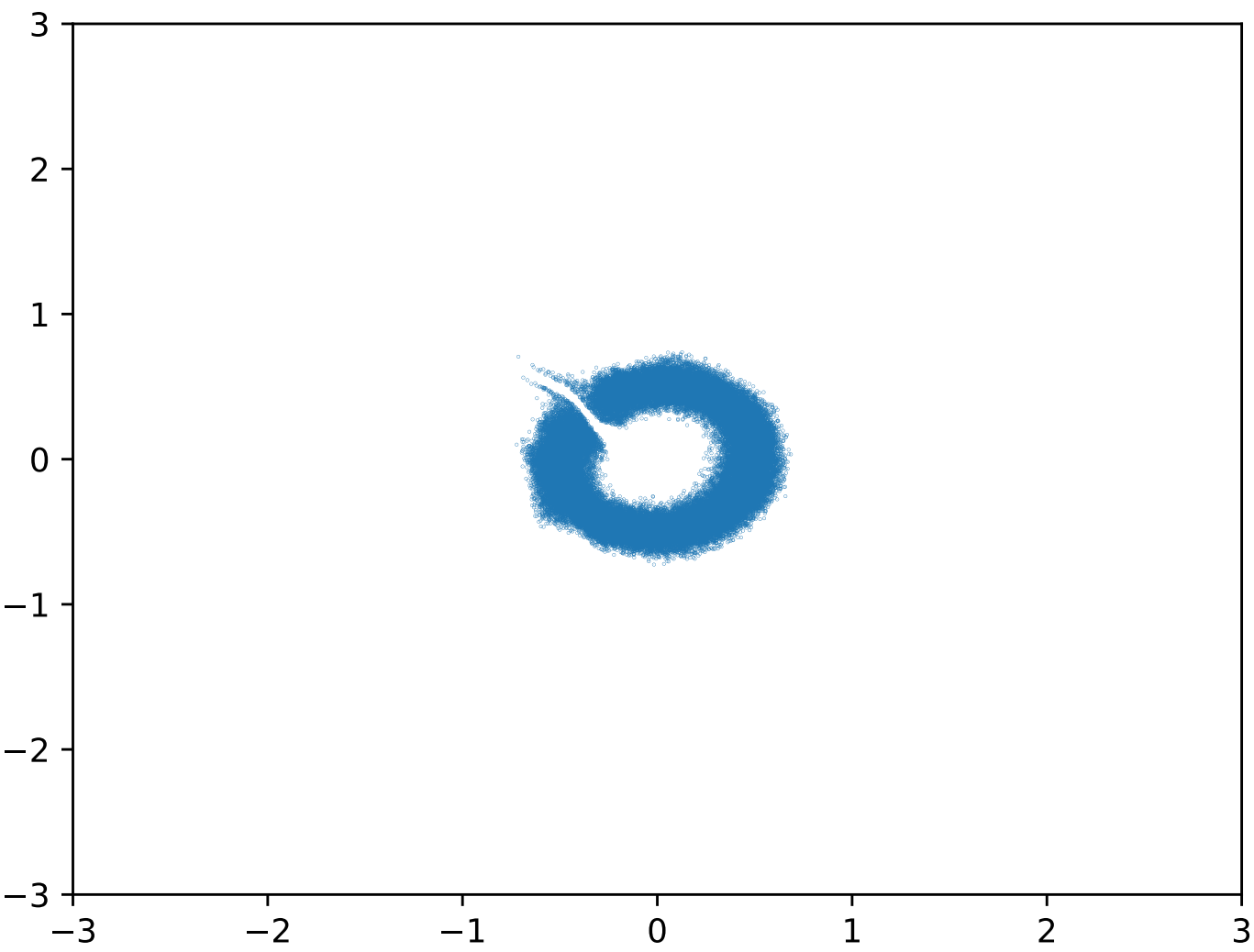}}
\;\;\;\;\;\;\;\;
\subfigure[Distribution generated using an unsupervised diffusion model]{\label{Diffusionring}
\includegraphics[width=0.4\textwidth]{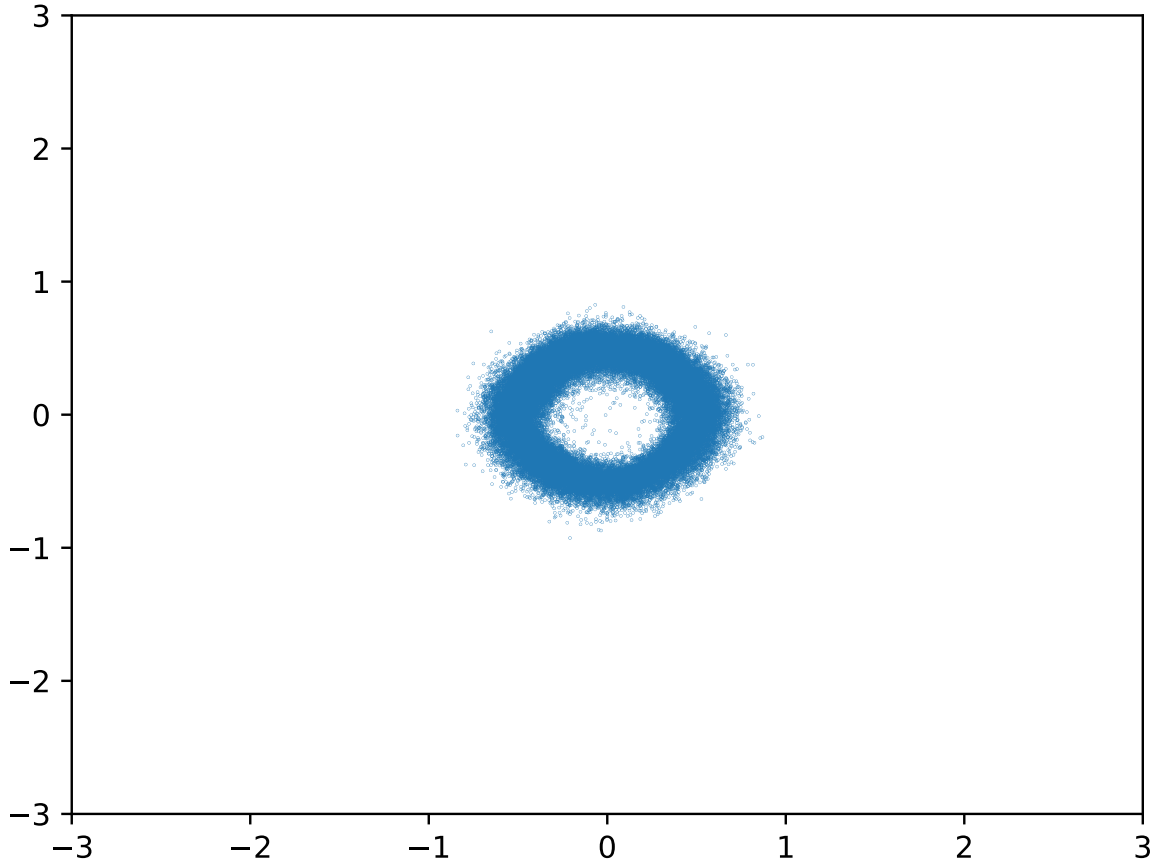}}
\caption{Target and prior distributions (above) and incomplete distributions generated using normalizing flow and an unsupervised diffusion model (below).
}
\label{manifold1}
\end{figure} 

The target, prior and output manifolds are shown as Fig.~\ref{manifold1}.
As shown in Fig.~\ref{Target}, the target manifold is made of three concentric circles and each of them forms a connected area. Fig.~\ref{Prior} demonstrates that the prior is a normal distribution with trivial topology. Given that homeomorphism maps preserve the topology, the output manifold from NFs is an open ring, which has trivial topology. This is consistent with the structure of the prior. The output manifold from the unsupervised DM is a closed ring (see~Fig.~\ref{Diffusionring}), in agreement with the discussion above. However, both unsupervised DMs and NFs here only produce inner ring configurations. The selection of this particular ring could be due to a higher density of the prior distribution near the origin.


\section{Stochastic pathway}

The major barrier to sampling configurations via NFs and unsupervised DMs for the outer rings is topology. Both NFs and unsupervised DMs keep the connectivity of the prior, despite the presence of white noise in the Langevin process. This is a major difference for the unsupervised DMs learning from the action, compared with the ones learning from a training data set. To sample configurations from a disconnected manifold, a randomized pathway is required.

\begin{figure}[thbp]
\centering
\includegraphics[width=0.80\textwidth]{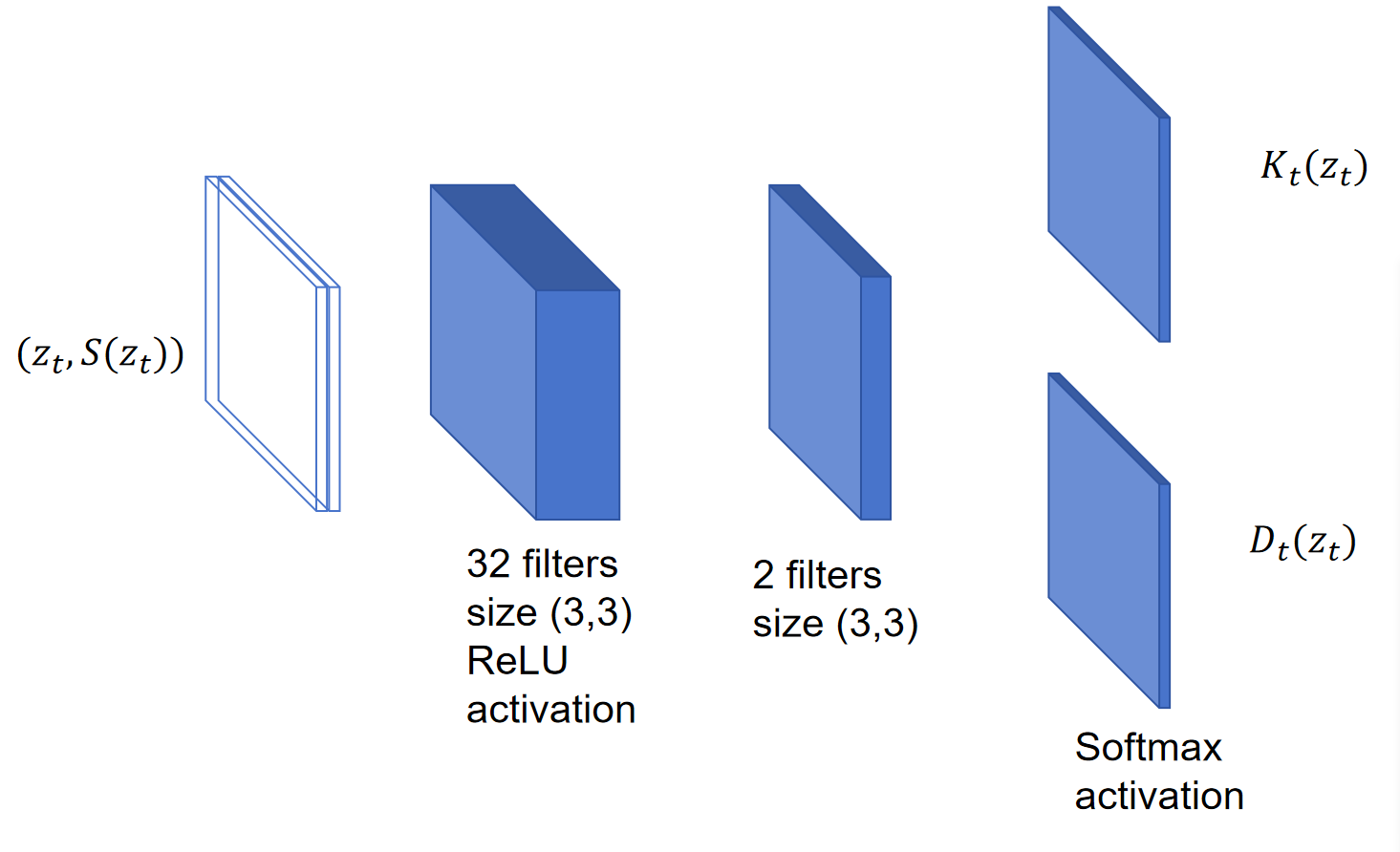}
\caption{Improved architecture for the forward and backward processes at fictitious time $t$.}
\label{newdiffusionnetwork}
\end{figure}

To improve sampling, the action $S(z_t) \propto \log P(z_t)$ should be included in neural network -- see Fig.~\ref{newdiffusionnetwork} -- since this may not be a continuous function.
Moreover, in addition to the state proposed by the forward process $P_{f,t}(z_{f,t}|z_{t},S(z_{t}))$, the state $z_{b,t}$ suggested by the backward process $P_{b,t}(z_{b,t}|z_{t},S(z_{t}))$ must also be considered, see~Fig.~\ref{diffusionnetwork} below.

\begin{figure}[thbp]
\centering
\includegraphics[width=0.80\textwidth]{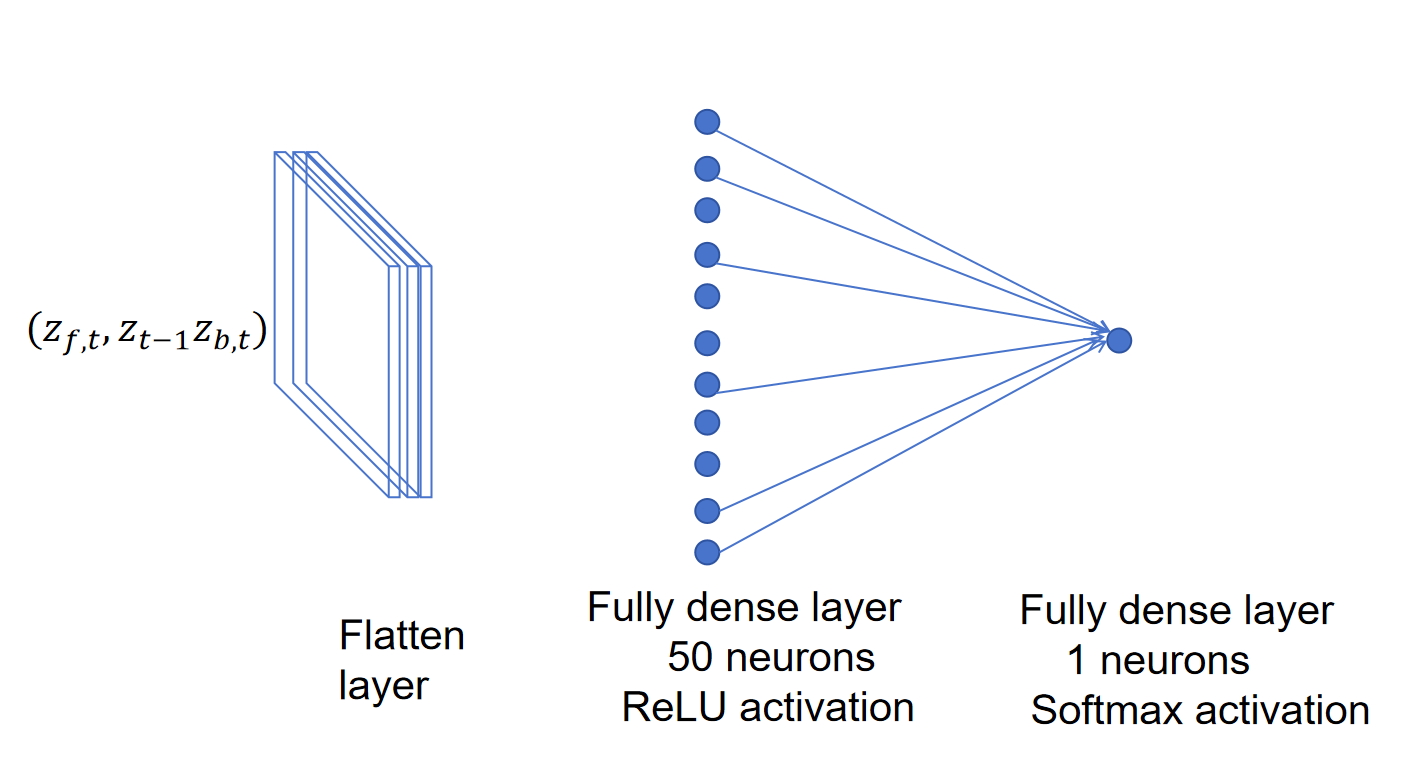}
\caption{The architecture for $P(z_t|z_{f,t-1},z_t,z_{b,t-1})$ and $P(z_{t-1}|z_{b,t},z_t,z_{f,t-1})$  at fictitious time $t$.}
\label{extrann}
\end{figure}

One more neural net $P(z_{f,t}, z_t,z_{b,t})$ with the proposed states $z_{f,t}$ and $z_{b,t}$ from the forward and backward processes is introduced to decide which pathways will be accepted. The network flattens the augmented pair $z_{f,t}$ and $z_{b,t}$, and then computes the probability of the upper pathway via two dense layers with 50 neurons and $\mathbf{ReLU}$ and $\mathbf{Sigmoid}$ activation functions, see Fig.~\ref{extrann}. The output probability is limited and less than $0.5$ to prevent pathway collapse.
The structure of the neural network at fictitious time $t$ is shown in Fig.~\ref{diffusionnetwork}. 
To optimize the network's parameters, the analytical formula for $P(z_t)$ is needed. This is given by

\begin{align}
    \frac{P(z_t)}{P(z_{t-1})} & =  \delta(z_t-z_{f,t})\frac{P(z_t|z_{f,t-1},z_t, z_{b,t-1})P_{f,t-1}(z_{f,t}|z_{t-1},S(z_{t-1}))}{P(z_{t-1}|z_{b,t},z_t,z_{f,t})P_{b,t}(z_{t-1}|z_{f,t},S(z_{f,t}))}
    \nonumber \\
    & + \delta(z_t-z_{b,t})\frac{P(z_t|z_{f,t-1},z_t, z_{b,t-1})P_{b,t-1}(z_{b,t}|z_{t-1},S(z_{t-1}))}{P(z_{t-1}|z_{b,t},z_t,z_{f,t})P_{f,t-1}(z_{t-1}|z_{b,t},S(z_{b,t-1}))}.
    \label{P(z_t)}
\end{align}
The first term of Eq.~\eqref{P(z_t)} means the state $z_{f,t}$ proposed by the forward process is accepted and the corresponding posterior $P_{b,t}(z_{t-1}|z_{t},S(z_{t}))$ is given by the backward process. Using Bayes' theorem, the left hand side of Eq.~\eqref{P(z_t)} is given the division of $P_{f,t-1}(z_{f,t}|z_{t-1},S(z_{t-1}))$ by $P_{b,t}(z_{b,t-1},|z_{t}, S(z_{t}))$, multiplying the corresponding weight. The weight is estimated by an extra network and is the probability of the forward pathway divided by $P(z_{t-1}|z_{b.t},z_t,z_{f,t})$, i.e.\ the probability of the forward pathway being selected using the same network. Similarly, one can compute the contribution from the backward process, which is shown as the second term in Eq.~\eqref{P(z_t)}. Combining Eqs.~\eqref{kl_divergence} and \eqref{P(z_t)} enables us to estimated the cost function and optimize the network.

\begin{figure}[thbp]
\centering
\includegraphics[width=0.90\textwidth]{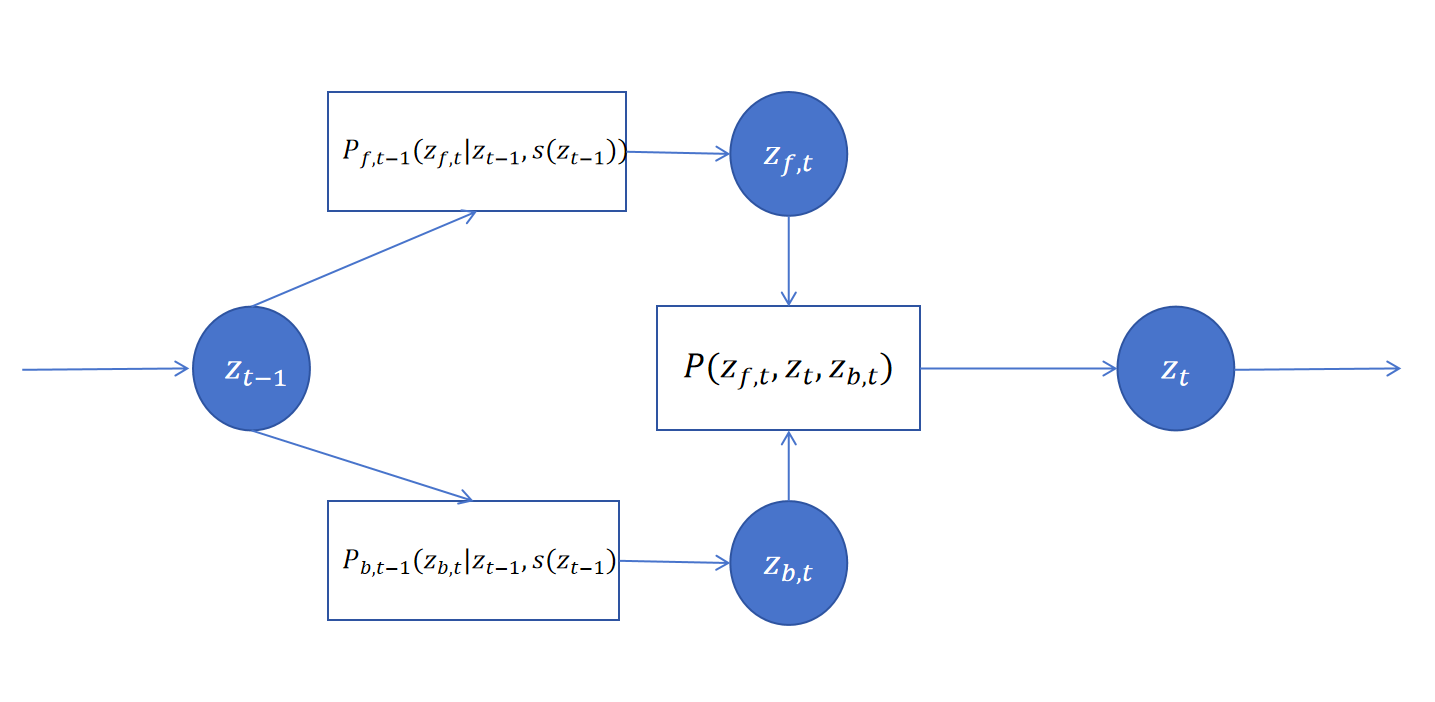}
\caption{The structure of the novel network at diffusion time $t$.}
\label{diffusionnetwork}
\end{figure}

\begin{figure}[htbp]
\centering
\includegraphics[width=0.45\textwidth]{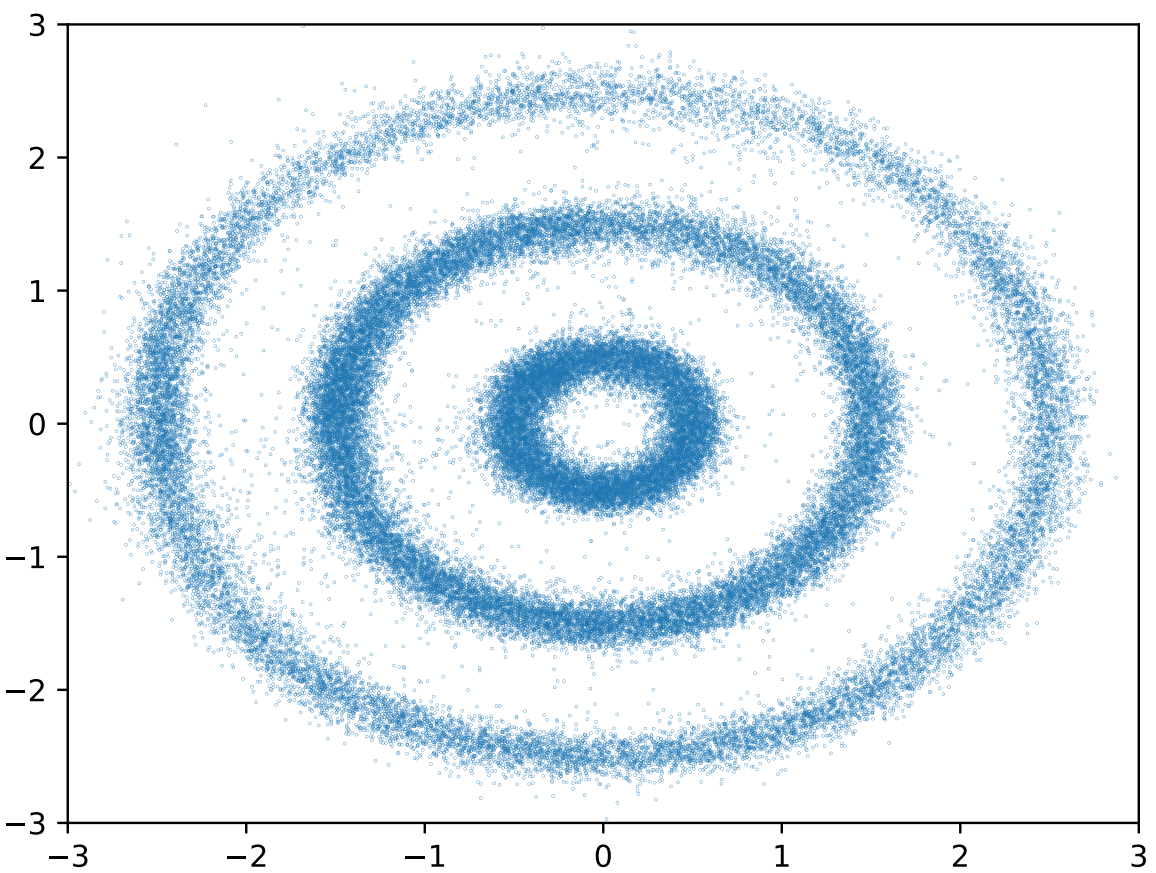}
\caption{Distribution generated using our novel neural net.}
\label{ourdiffusion}
\end{figure}

 The output manifold from our modified neural is shown in Fig.~\ref{ourdiffusion}. Our network reproduces the features of the target distribution. We note however that the samples are noisy. This happens because for a diffusion model, the drift of the backward process at fictitious time $i$ includes $\nabla \log P(z_i)$. In our approach, this term is estimated by the neural network $P_{b,i}(z_i|z_{i-1}, S(z_{i-1}))$. 
 
\section{Scalar field theory}

\begin{figure}[thbp]
\centering
\subfigure{
\label{tworingNF}
\includegraphics[width=0.7\textwidth]{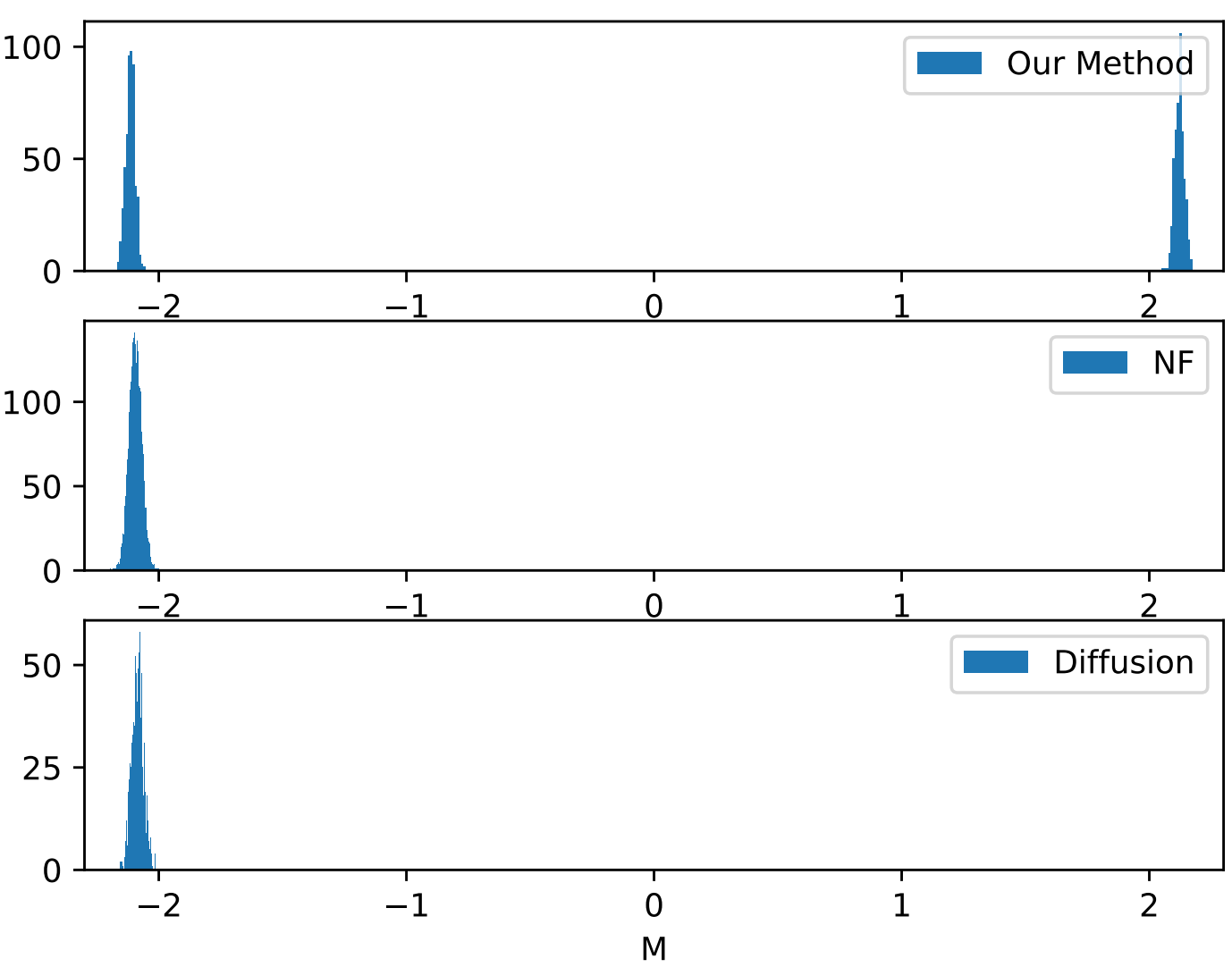}}
\caption{Histograms of the magnetisation $M$ in the two-dimensional scalar field theory obtained using the novel architecture introduced in this work (above), normalising flow (middle) and unsupervised diffusion models (below).}
\label{phi4results}
\end{figure}

The physics model we used here is a self-interacting real scalar field on a two-dimensional ($d=2$) lattice, $|\Lambda|=N_{T}\times N_{L}$. The corresponding action is
\begin{equation}
S(\phi) = \sum\limits_{x\in \Lambda} \left(-2\kappa\sum\limits_{\hat{\mu}=1}^{d}\phi(x)\phi(x+\hat{\mu})+(1-2\lambda)\phi^2(x) + \lambda \phi^4(x)\right),
\label{action}
\end{equation}
where $\kappa$ is the hopping parameter and $\lambda$ the bare coupling constant. 
The model has a $Z_2$ symmetry. The symmetry can be observed by estimating the magnetization,
\begin{equation}
	M = \frac{1}{N_T N_L}\sum\limits_{x\in \Lambda}\phi_{x}.
\end{equation}
 We have estimated the magnetization for the parameter values $\lambda=0.022$ and $\kappa=0.3$ at a volume $|\Lambda|=64\times32$, comparing results obtained with our novel approach, normalising flow, and the unsupervised diffusion model. 
 Histograms of $M$ are shown in Fig.~\ref{phi4results}. This figure demonstrates that NFs and unsupervised DMs recreate one peak, whereas our novel network generates the correct distribution with two peaks. 

\section{Summary}

In this contribution, we have studied the performance of NFs and unsupervised DMs in a triple-ring model. Since the prior manifold has a trivial topology and the target distribution has three disconnected components, both NFs and unsupervised DMs suffer from ergodicity problems connected with the presence of non-trivial topological sectors, leading to some regions in configuration space not being generated. To sample data from the correct support set, in each diffusion step, $z_{b,t}$ from the backward network of unsupervised DMs can be used. One more network is introduced to initiate a route by choosing between $z_{b,t}$ and $z_{f,t}$ from the forward network. We find that this new structure of the network is able to randomize the pathway so as to improve sampling for the triple-ring model. In two-dimensional $\phi^4$ model, NFs and unsupervised DMs reproduce only a single peak for the magnetization, while our method results in configurations obeying the expected $Z_2$ symmetry.

\acknowledgments

SYC is supported by the China Scholarship Council (No.~202308420042) and a Swansea University joint PhD project. 
GA and BL are supported by STFC Consolidated Grant ST/X000648/1.  
BL is further supported by the UKRI EPSRC ExCALIBUR ExaTEPP project EP/X017168/1.

\noindent
{\bf Research Data and Code Access} --
The code and data used for in this manuscript is available upon request.

\bibliographystyle{JHEP}
\bibliography{refs}

\providecommand{\href}[2]{#2}\begingroup\raggedright\begin{thebibliography}{10}

\bibitem{Wolff:1989wq}
U.~Wolff, \emph{{CRITICAL SLOWING DOWN}},
  \href{https://doi.org/10.1016/0920-5632(90)90224-I}{\emph{Nucl. Phys. B Proc.
  Suppl.} {\bfseries 17} (1990) 93}.

\bibitem{DelDebbio:2004xh}
L.~Del~Debbio, G.M.~Manca and E.~Vicari, \emph{{Critical slowing down of
  topological modes}},
  \href{https://doi.org/10.1016/j.physletb.2004.05.038}{\emph{Phys. Lett. B}
  {\bfseries 594} (2004) 315}
  [\href{https://arxiv.org/abs/hep-lat/0403001}{{\ttfamily hep-lat/0403001}}].

\bibitem{Schaefer:2010hu}
{\scshape ALPHA} collaboration, \emph{{Critical slowing down and error analysis
  in lattice QCD simulations}},
  \href{https://doi.org/10.1016/j.nuclphysb.2010.11.020}{\emph{Nucl. Phys. B}
  {\bfseries 845} (2011) 93} [\href{https://arxiv.org/abs/1009.5228}{{\ttfamily
  1009.5228}}].

\bibitem{Cranmer:2023xbe}
K.~Cranmer, G.~Kanwar, S.~Racani\`ere, D.J.~Rezende and P.E.~Shanahan,
  \emph{{Advances in machine-learning-based sampling motivated by lattice
  quantum chromodynamics}},
  \href{https://doi.org/10.1038/s42254-023-00616-w}{\emph{Nature Rev. Phys.}
  {\bfseries 5} (2023) 526} [\href{https://arxiv.org/abs/2309.01156}{{\ttfamily
  2309.01156}}].

\bibitem{Kanwar:2024ujc}
G.~Kanwar, \emph{{Flow-based sampling for lattice field theories}},
  \href{https://doi.org/10.22323/1.453.0114}{\emph{PoS} {\bfseries LATTICE2023}
  (2024) 114} [\href{https://arxiv.org/abs/2401.01297}{{\ttfamily
  2401.01297}}].

\bibitem{Aarts:2025gyp}
G.~Aarts, K.~Fukushima, T.~Hatsuda, A.~Ipp, S.~Shi, L.~Wang et~al.,
  \emph{{Physics-driven learning for inverse problems in quantum
  chromodynamics}},
  \href{https://doi.org/10.1038/s42254-024-00798-x}{\emph{Nature Rev. Phys.}
  (2025) } [\href{https://arxiv.org/abs/2501.05580}{{\ttfamily 2501.05580}}].

\bibitem{Pawlowski:2018qxs}
J.M.~Pawlowski and J.M.~Urban, \emph{{Reducing Autocorrelation Times in Lattice
  Simulations with Generative Adversarial Networks}},
  \href{https://doi.org/10.1088/2632-2153/abae73}{\emph{Mach. Learn. Sci.
  Tech.} {\bfseries 1} (2020) 045011}
  [\href{https://arxiv.org/abs/1811.03533}{{\ttfamily 1811.03533}}].

\bibitem{Wang:2020hji}
L.~Wang, Y.~Jiang, L.~He and K.~Zhou, \emph{{Continuous-Mixture Autoregressive
  Networks Learning the Kosterlitz-Thouless Transition}},
  \href{https://doi.org/10.1088/0256-307X/39/12/120502}{\emph{Chin. Phys.
  Lett.} {\bfseries 39} (2022) 120502}
  [\href{https://arxiv.org/abs/2005.04857}{{\ttfamily 2005.04857}}].

\bibitem{Wang:2023exq}
L.~Wang, G.~Aarts and K.~Zhou, \emph{{Diffusion models as stochastic
  quantization in lattice field theory}},
  \href{https://doi.org/10.1007/JHEP05(2024)060}{\emph{JHEP} {\bfseries 05}
  (2024) 060} [\href{https://arxiv.org/abs/2309.17082}{{\ttfamily
  2309.17082}}].

\bibitem{Wang:2023sry}
L.~Wang, G.~Aarts and K.~Zhou, \emph{{Generative Diffusion Models for Lattice
  Field Theory}},  in \emph{{37th Conference on Neural Information Processing
  Systems}}, 2023 [\href{https://arxiv.org/abs/2311.03578}{{\ttfamily
  2311.03578}}].

\bibitem{Zhu:2024kiu}
Q.~Zhu, G.~Aarts, W.~Wang, K.~Zhou and L.~Wang, \emph{{Diffusion models for
  lattice gauge field simulations}},  in \emph{{38th conference on Neural
  Information Processing Systems}}, 2024
  [\href{https://arxiv.org/abs/2410.19602}{{\ttfamily 2410.19602}}].

\bibitem{Aarts:2024rsl}
G.~Aarts, D.E.~Habibi, L.~Wang and K.~Zhou, \emph{{On learning higher-order
  cumulants in diffusion models}},  in \emph{{38th conference on Neural
  Information Processing Systems}}, 2024
  [\href{https://arxiv.org/abs/2410.21212}{{\ttfamily 2410.21212}}].

\bibitem{rezende2015variational}
D.J.~Rezende and S.~Mohamed, \emph{{Variational Inference with Normalizing
  Flows}},  in \emph{International conference on machine learning},
  pp.~1530--1538, PMLR, 2015
  [\href{https://arxiv.org/abs/1505.05770}{{\ttfamily 1505.05770}}].

\bibitem{Noe:2019}
F.~Noé, S.~Olsson, J.~Köhler and H.~Wu, \emph{Boltzmann generators: Sampling
  equilibrium states of many-body systems with deep learning},
  \href{https://doi.org/10.1126/science.aaw1147}{\emph{Science} {\bfseries 365}
  (2019) eaaw1147} [\href{https://arxiv.org/abs/1812.01729}{{\ttfamily
  1812.01729}}].

\bibitem{Albergo:2019eim}
M.S.~Albergo, G.~Kanwar and P.E.~Shanahan, \emph{Flow-based generative models
  for {{Markov}} chain {{Monte Carlo}} in lattice field theory},
  \href{https://doi.org/10.1103/PhysRevD.100.034515}{\emph{Phys. Rev. D}
  {\bfseries 100} (2019) 034515}
  [\href{https://arxiv.org/abs/1904.12072}{{\ttfamily 1904.12072}}].

\bibitem{Nicoli:2019gun}
K.A.~Nicoli, S.~Nakajima, N.~Strodthoff, W.~Samek, K.-R.~M\"uller and
  P.~Kessel, \emph{{Asymptotically unbiased estimation of physical observables
  with neural samplers}},
  \href{https://doi.org/10.1103/PhysRevE.101.023304}{\emph{Phys. Rev. E}
  {\bfseries 101} (2020) 023304}
  [\href{https://arxiv.org/abs/1910.13496}{{\ttfamily 1910.13496}}].

\bibitem{Kanwar:2020xzo}
G.~Kanwar, M.S.~Albergo, D.~Boyda, K.~Cranmer, D.C.~Hackett, S.~Racani{\`e}re
  et~al., \emph{Equivariant {{Flow-Based Sampling}} for {{Lattice Gauge
  Theory}}}, \href{https://doi.org/10.1103/PhysRevLett.125.121601}{\emph{Phys.
  Rev. Lett.} {\bfseries 125} (2020) 121601}
  [\href{https://arxiv.org/abs/2003.06413}{{\ttfamily 2003.06413}}].

\bibitem{Nicoli:2020njz}
K.A.~Nicoli, C.J.~Anders, L.~Funcke, T.~Hartung, K.~Jansen, P.~Kessel et~al.,
  \emph{Estimation of {{Thermodynamic Observables}} in {{Lattice Field
  Theories}} with {{Deep Generative Models}}},
  \href{https://doi.org/10.1103/PhysRevLett.126.032001}{\emph{Phys. Rev. Lett.}
  {\bfseries 126} (2021) 032001}
  [\href{https://arxiv.org/abs/2007.07115}{{\ttfamily 2007.07115}}].

\bibitem{Nicoli:2023qsl}
K.A.~Nicoli, C.J.~Anders, T.~Hartung, K.~Jansen, P.~Kessel and S.~Nakajima,
  \emph{{Detecting and mitigating mode-collapse for flow-based sampling of
  lattice field theories}},
  \href{https://doi.org/10.1103/PhysRevD.108.114501}{\emph{Phys. Rev. D}
  {\bfseries 108} (2023) 114501}
  [\href{https://arxiv.org/abs/2302.14082}{{\ttfamily 2302.14082}}].

\bibitem{Chen:2018}
R.T.~Chen, Y.~Rubanova, J.~Bettencourt and D.K.~Duvenaud, \emph{Neural ordinary
  differential equations}, {\emph{Advances in neural information processing
  systems} {\bfseries 31} (2018) }
  [\href{https://arxiv.org/abs/1806.07366}{{\ttfamily 1806.07366}}].

\bibitem{deHaan:2021erb}
P.~de~Haan, C.~Rainone, M.C.N.~Cheng and R.~Bondesan, \emph{{Scaling Up Machine
  Learning For Quantum Field Theory with Equivariant Continuous Flows}},
  \href{https://arxiv.org/abs/2110.02673}{{\ttfamily 2110.02673}}.

\bibitem{Gerdes:2022eve}
M.~Gerdes, P.~de~Haan, C.~Rainone, R.~Bondesan and M.C.N.~Cheng,
  \emph{{Learning lattice quantum field theories with equivariant continuous
  flows}}, \href{https://doi.org/10.21468/SciPostPhys.15.6.238}{\emph{SciPost
  Phys.} {\bfseries 15} (2023) 238}
  [\href{https://arxiv.org/abs/2207.00283}{{\ttfamily 2207.00283}}].

\bibitem{Caselle:2023mvh}
M.~Caselle, E.~Cellini and A.~Nada, \emph{{Sampling the lattice Nambu-Goto
  string using Continuous Normalizing Flows}}, {\emph{Journal of High Energy
  Physics} {\bfseries 02} (2024) 048}
  [\href{https://arxiv.org/abs/2307.01107}{{\ttfamily 2307.01107}}].

\bibitem{wu2020stochastic}
H.~Wu, J.~K{\"o}hler and F.~No{\'e}, \emph{Stochastic normalizing flows},
  {\emph{Advances in Neural Information Processing Systems} {\bfseries 33}
  (2020) 5933} [\href{https://arxiv.org/abs/2002.06707}{{\ttfamily
  2002.06707}}].

\bibitem{Caselle:2022acb}
M.~Caselle, E.~Cellini, A.~Nada and M.~Panero, \emph{Stochastic normalizing
  flows as non-equilibrium transformations},
  \href{https://doi.org/10.1007/JHEP07(2022)015}{\emph{JHEP} {\bfseries 07}
  (2022) 015} [\href{https://arxiv.org/abs/2201.08862}{{\ttfamily
  2201.08862}}].

\bibitem{Caselle:2024ent}
M.~Caselle, E.~Cellini and A.~Nada, \emph{{Numerical determination of the width
  and shape of the effective string using Stochastic Normalizing Flows}},
  \href{https://arxiv.org/abs/2409.15937}{{\ttfamily 2409.15937}}.

\bibitem{DBLP:journals/corr/abs-2106-04399}
E.~Bengio, M.~Jain, M.~Korablyov, D.~Precup and Y.~Bengio, \emph{Flow network
  based generative models for non-iterative diverse candidate generation},
  \href{https://arxiv.org/abs/2106.04399}{{\ttfamily 2106.04399}}.

\bibitem{DBLP:journals/corr/abs-2111-09266}
Y.~Bengio, T.~Deleu, E.J.~Hu, S.~Lahlou, M.~Tiwari and E.~Bengio,
  \emph{Gflownet foundations},
  \href{https://arxiv.org/abs/2111.09266}{{\ttfamily 2111.09266}}.

\bibitem{Albergo:2021vyo}
M.S.~Albergo, D.~Boyda, D.C.~Hackett, G.~Kanwar, K.~Cranmer, S.~Racani\`ere
  et~al., \emph{{Introduction to Normalizing Flows for Lattice Field Theory}},
  \href{https://arxiv.org/abs/2101.08176}{{\ttfamily 2101.08176}}.

\end{thebibliography}\endgroup

\end{document}